\documentclass[11pt,twoside]{article}

\usepackage{asp2004}
\usepackage{epsf}
\usepackage{psfig}
\usepackage{lscape}

\begin{document}
\title{Galaxy Formation}   
\author{Eric Gawiser}   
\affil{NSF Astronomy \& Astrophysics Postdoctoral Fellow (AAPF), Yale 
Astronomy Department and Yale Center for Astronomy \& Astrophysics, 
PO Box 208101, New Haven, CT 06520-8101}    

\begin{abstract} 
I summarize current knowledge of galaxy formation 
with emphasis on the initial conditions 
provided by the $\Lambda$CDM cosmology, integral constraints 
from cosmological quantities, and the demographics of high-redshift 
protogalaxies.  
Tables are provided summarizing the number density, star formation 
rate and stellar mass per object, cosmic star formation rate and 
stellar mass densities, clustering length and typical 
dark matter halo masses for Lyman break galaxies, Lyman alpha 
emitting galaxies, Distant red galaxies, Sub-millimeter galaxies, 
and Damped Lyman $\alpha$ absorption systems.  
I also discuss five key unsolved problems 
in galaxy formation and prognosticate advances that the near 
future will bring.
\end{abstract}



\section{Boundary Conditions for Galaxy Formation}
\label{sec:boundary}

\subsection{Initial Conditions:  \boldmath $\Lambda$CDM Cosmology}
\label{sec:initial}

The initial conditions for the formation of galaxies are provided 
by the now-standard $\Lambda$CDM cosmological model.  
The combined results of the WMAP satellite study of 
Cosmic Microwave Background anisotropies, large-scale 
structure, and Type Ia supernovae observations yield 
best-fit values for the cosmological parameters of 
roughly $\Omega_\Lambda = 0.7$, 
$\Omega_m=0.3$, $\Omega_b = 0.04$, and 
$H_0 = 70h_{70}$km s$^{-1}$ Mpc$^{-1}$ 
\citep{bennettetal03}.\footnote{We include 
$h_{70}$, analogous to the traditional parameter $h\equiv h_{100}$, 
even though 
its value appears quite close to 1.}  
The original model of galaxy formation was  
Monolithic Collapse \citep*{eggenls62}, 
where gravitational collapse of a cloud of primordial gas 
very early in the lifetime of the Universe formed all parts of 
each galaxy at the same time.  Modern evidence rules out this 
model on two fronts; the widely varying ages of different 
components of the Galaxy provide a counter-example, and the $\Lambda$CDM 
cosmology predicts ``bottom-up'' i.e. hierarchical rather 
than ``top-down'' structure formation.  


Hierarchical structure formation is a generic feature of Cold 
Dark Matter (CDM) models.  Small overdensities are able to 
overcome the cosmological expansion and collapse first, 
and the resulting dark matter ``halos'' merge together to form 
larger halos which serve as sites of galaxy formation.
This process continues until the present day, 
making galaxy formation an ongoing process.
The nearly-scale-invariant primordial power spectrum inferred 
from combining WMAP with large-scale structure observations provides 
power on all scales in the distribution of CDM.  
The  
baryons fall into the CDM potential wells after decoupling, leaving 
only trace evidence of their previous acoustic oscillations as a series of 
low-amplitude peaks in the matter power spectrum.
The non-linear collapse of dark matter overdensities 
occurs on larger and larger scales, so the typical collapsed halo 
mass grows with time, but no preferred scale is introduced.  
$\Lambda$CDM therefore provides a distribution of halos where galaxies 
can form, with the details of the process up to baryonic physics.  

Despite the lack of preferred galaxy scales in the distribution of 
dark matter halos, baryonic physics causes galaxies to 
have minimum and maximum 
masses.  
The maximum mass is that of CD galaxies in cluster centers 
with baryonic masses $\sim 10^{12}\mathrm{M}_\odot$ and 
virial masses  $\sim 10^{13}\mathrm{M}_\odot$; there are 
$\sim10^{14}$M$_\odot$ of baryons available in a rich cluster but 
virialization of galaxies and heating of gas to the high virial temperature  
prevent most of this mass 
from finding its way to the central galaxy.  The minimum mass observed 
today is that of dwarf galaxies, $\sim10^8\mathrm{M}_\odot$, but galaxies 
may initially 
have formed as small as  $10^6\mathrm{M}_\odot$ (the baryonic Jean's mass after recombination 
i.e. the minimum mass for which gravity overwhelmed 
pressure support).  Explaining the lack of observed galaxies  
with circular velocities below 30 km/s is a major goal; it is suspected 
that feedback from supernovae explosions may have quenched star formation in 
such low-mass objects immediately after a single burst of star formation 
\citep{dekels86}.  

The growth of cosmological structure and collapse of dark matter halos 
is a feature of the matter-dominated epoch.  During radiation-domination, 
perturbations on scales smaller than the sound horizon were unable to 
grow due to acoustic oscillations in the photon-baryon fluid that gave 
rise to the famous peaks in the CMB angular power spectrum and the 
lower-amplitude peaks in the matter power spectrum.  Now that we have entered 
a phase of dark energy domination, structure growth is slowing and will 
cease entirely as the universe enters a new phase of inflation.  This 
cosmological ``freeze-out'' in structure formation is recent, since 
equality between the dark energy and matter densities occurred 
at $z_{eq}=0.4$. The slowing of structure formation occurs gradually, 
so the growth of cosmological structure 
continued nearly unabated until $z_{eq}$, even though we see strong 
observational evidence for ``downsizing'' at $z<1$ where high-mass 
galaxies grow far more slowly than lower-mass galaxies 
\citep[e.g.][]{treuetal05,smith05}.  
Another term being used by some is ``anti-hierarchical'', 
which is basically a synonym for ``downsizing'' but seems to 
imply inconsistency with hierarchical cosmology.  However, the observed 
freeze-out in galaxy (and possibly 
supermassive black hole) formation in massive galaxies is 
not inconsistent with CDM models; rather, 
it appears to be 
caused by baryonic feedback which is not well understood at 
present (see \S \ref{sec:problems}).  
The slowing of cosmological 
structure growth since $z\simeq0.4$ may, 
however, play a role in the recent decline 
of the cosmic star formation rate density discussed by \citet{belletal05}.  

\subsection{Final Conditions:  Low-redshift Galaxies}
\label{sec:lowz}
The study of galaxy formation is made easier by having full 
boundary conditions.  The final conditions are the Hubble sequence 
of mature galaxies we see in the nearby universe at redshift zero.  Indeed, 
much has been learned about galaxy formation from ``archaeological'' 
evidence in the ages and chemical abundances of various Galactic 
stellar populations, and expanding these studies to the rest of 
the Local Group and beyond is quite useful.  Nonetheless, there are 
great advantages to observing galaxies in the act of formation, 
which motivates the study of high-redshift galaxies.  
At $z>2$, 
galaxy-mass halos are rare so the majority of galaxies we observe 
reside in dark matter halos that have only recently 
collapsed i.e. at high-redshift most galaxies are young.  
In this sense, $z>2$ can be considered the epoch of galaxy 
formation.

\section{Integral Constraints:  Cosmological Quantities}
\label{sec:ic}

Instead of studying galaxies as discrete objects residing in 
dark matter halos, one can track the cosmological quantities that 
comprise the baryon budget.  Galaxy formation and evolution plays 
the fundamental role in the processing of baryons from neutral 
hydrogen to molecular gas to stars to metals.   
Star formation is inextricably linked with galaxy formation; whether 
you choose to define a galaxy as a large conglomeration of stars or an 
overdensity of baryons inside a collapsed dark matter halo, the galaxies 
in our universe form great numbers of stars.
The cosmological quantities of interest provide integral constraints 
on star formation.  The cosmic star formation rate density 
(SFRD) is an integral constraint averaged over the volume of the 
universe observable at a given redshift.  
The cosmic density of 
neutral gas, $\Omega_{gas}$, the cosmic density of metals, $\Omega_Z$, and 
the cosmic stellar mass density all provide integral constraints on the 
SFRD over time, as will be discussed below.  
The sum of the cosmic infrared background 
(CIB) and cosmic far-infrared background (FIRB) radiation provides an 
integral constraint on the SFRD from the Big Bang all the way to 
$z=0$ by tracing the energy generated by nuclear reactions 
in stars.  

\subsection{Cosmic Density of Neutral Gas}
\label{sec:HI}

The Damped Lyman $\alpha$ Absorption systems (DLAs, 
\citealp{wolfeetal86}) are quasar absorption 
line systems with HI column densities $\geq 2\times10^{20}$cm$^{-2}$, 
sufficient to self-shield against the high-redshift ionizing 
background. 
Studying quasar absorption-line spectra provides a 
(nearly) unbiased sample of lines-of-sight through the cosmos 
ideal for measuring cosmological quantities.  
The DLAs have been found to 
contain the majority of neutral hydrogen atoms at 
high redshift 
\citep*[see the recent review by][]{wolfegp05}.
Moreover, 
DLAs contain the vast majority of neutral gas, by which 
we mean neutral hydrogen and helium in regions that are 
sufficiently neutral to cool and participate in star 
formation, as lower column density systems 
are predominantly ionized.  
  Hence the DLAs provide the reservoir of 
neutral gas that is available for star formation.  
In a simple closed box model, 
$d\rho_{gas}/dt = -d\rho_*/dt$, and 
the net decrease in the cosmic density 
of neutral gas from $z=3$ to $z=0$ 
is assumed to have all been turned into stars
\citep[see Fig. 5 of ][]{wolfegp05}.
In that case, the 
DLAs appear to have formed about half of the stars seen in 
galaxies today.  The truth is more complicated in 
hierarchical cosmology, where an open box model must be used;   
\begin{equation}
\frac{d\rho_{gas}}{dt} = -\frac{d\rho_*}{dt} 
    + \mathrm{infall} + \mathrm{merging} - \mathrm{winds}  . \; \; \; 
\label{eq:openbox}
\end{equation} 
Cosmological models for infall of gas from the intergalactic 
medium (IGM), merging of lower column-density systems, 
and gas loss due to galactic winds are still quite 
uncertain, but the star formation rates actually measured for 
DLAs \citep*[][\citealp{wolfeetal04}]{wolfegp03,wolfepg03} 
imply that DLAs could have formed 
all present-day stars.  Unfortunately, 
large uncertainties in the source and sink terms prevent us from using  
changes in the cosmic density of neutral gas as an integral constraint 
on the cosmic SFRD at the present time.  

\subsection{Star Formation Rate Density}
\label{sec:sfrd}

The cosmic star formation rate density has now been measured out 
to $z\simeq6$
\citep{giavaliscoetal04}.
The high-redshift points are taken from only the Lyman break galaxies, and 
it is unclear how severe the resulting incompleteness is 
since we are not sure if all star-forming galaxy populations at these 
redshifts are known.  The plot is traditionally shown in 
misleading units of 
M$_\odot$Mpc$^{-3}$yr$^{-1}$ versus redshift;
in order to integrate-by-eye, one should plot 
this quantity versus time, and this 
has the effect of greatly increasing the apparent amount of star formation 
at low redshifts.  Despite significant uncertainties in the SFRD 
at $z>3$ due to incompleteness and large dust corrections, 
it appears that most stars in the present-day 
universe formed at $z<2$ \citep[see Fig. 33 of][]{pettini04}.  

\subsection{Stellar Mass Density}
\label{sec:stellarmass}

The cosmic stellar mass density provides an integral 
constraint on the SFRD, $\rho_*(t)=\int_0^t d\rho_*/dt$.  
See \citet{dickinsonetal03} for a recent compilation, and 
Niv Drory's contribution to this volume for an update.   
Note that the stellar masses of galaxies are not direct observables 
but are inferred from rest-frame optical (and near-infrared) photometry 
by modelling each object's star formation history using an assumed 
initial mass function (IMF).

\subsection{Cosmic Metal Enrichment History}
\label{sec:metals}
The cosmic metal density is really a history of 
cosmic metal enrichment due to star formation,                  
$\rho_*(t)=1/42 \int_0^t d\rho_*/dt$ \citep{pettini04}.      
\citeauthor{wolfegp05} (\citeyear{wolfegp05}, see their Fig. 7) 
show that the 
cosmic metallicity traced by DLAs rises gradually 
from  
a mean value of [M/H]=-1.5 at $z\simeq 4$ 
to a mean value of -0.7 at $z\simeq 1$.  
The range of observed DLA metallicities is somewhat higher than 
that of halo stars but overlaps, and is somewhat lower than that of thick disk 
stars and far lower than the near-solar values seen for thin disk 
stars in the Milky Way.  The DLAs uniformly show 
greater metal enrichment than the Lyman $\alpha$ forest 
but less than values inferred for Lyman break galaxies or quasars 
at the same epoch 
\citep[see Figs. 8, 32 of][and see \citealp{leitherer05} for a review]{pettini04}.  
The values given above are the cosmic mean metallicity of the neutral gas 
traced by DLAs, 
but they do not represent a full census of metals, which 
can also be found in heavily star-forming regions that have already 
used up their neutral gas or can be expelled by galactic winds 
into the IGM, which is predominantly ionized.  
It is therefore useful to compare the observed DLA metallicities 
with those expected from the DLA star formation rates; this leads 
to a factor of ten deficit in the observed metallicities called the 
``Missing Metals Problem'' 
\citep{wolfegp05,hopkinsrt05,pettini99}.  
  The most likely explanation is that 
the star-forming regions of the galaxies seen as DLAs have superwinds 
sufficiently strong to move most of the metals produced into the IGM.

\section{Theoretical Advances}
\label{sec:theory}  

Theoretical efforts to understand and model galaxy formation are 
mostly beyond the analytical realm, where they divide into semi-analytic  
models and cosmological simulations.  These two approaches have been 
converging in recent years, as the practitioners of cosmological 
hydrodynamic simulations are using more detailed ``recipes'' for 
star formation, supernova feedback, and winds and in some cases 
have claimed grandiose results from purely N-body simulations 
with many semi-analytic recipes added
\citep[e.g.][]{springeletal05}.  For examples of state-of-the-art 
cosmological hydrodynamic simulations of high-redshift galaxies 
and AGN, see \citet*{nagamineetal04a} and \citet*{dimatteoetal05}.  

Semi-analytic models reproduce observations moderately well but have 
yet to demonstrate much success in predicting future observations, 
making them more of a tool for interpreting results than theoretical 
models in the classic sense.  
\citet*{somervilleetal01} tuned their models 
to reproduce galaxy properties at $z=0$ and found one of their models 
to be in good agreement with the dust-corrected points at 
$z>2$ in the cosmic SFRD diagram.  
However, as mentioned above, semi-analytic models for infall, merging, and 
winds are highly uncertain and it is not clear if observations of 
the cosmic density of neutral gas and the missing metals problem 
are consistent with the predictions.  Similar scatter is seen in 
theoretical predictions of the cosmic stellar mass density.

\section{Protogalaxy Demographics}
\label{sec:census}  

DLAs dominate the neutral gas, making DLA-based studies appropriate 
for determining its cosmic density.  However,  other 
cosmological quantities should be summed over all high-redshift objects 
rather than just DLAs or just Lyman break galaxies, which trace the 
bright end of the high-redshift rest-UV galaxy luminosity function.  
Another motivation for studying all types of objects 
is the search for the progenitors of typical spiral 
galaxies like the Milky Way, which have not yet been pinpointed amongst 
the zoo of high-redshift galaxies.  In designing the 
Multiwavelength Survey by Yale-Chile (MUSYC, \citealp{gawiseretal05}, 
http://www.astro.yale.edu/MUSYC), 
it was decided to focus on selecting all known populations of 
galaxies at $z\simeq3$, where most objects are young and several 
selection techniques overlap
\citep[see review by][]{sterns99}.  The various populations at this 
epoch are  
labelled by three-letter acronyms (TLAs).  We discuss each below.  

\subsection{Lyman Break Galaxies (LBGs)}
\label{sec:lbg}

The Lyman break galaxies (LBGs) are 
   selected via the Lyman break 
at 912\AA\ in the rest-frame.  Higher-energy photons are 
unable to escape the galaxies or travel far in the IGM 
due to the large cross-section for absorption of ionizing 
photons by neutral hydrogen 
\citep[for an illustration of the technique first successfully 
applied by \citealp{steidelh92}, see Fig. 19 of][]{pettini04}.  
At $z\simeq 3$, the Lyman break generates a very red color in 
$U-V$, which could also be observed for an intrinsically red object 
such as an M dwarf or elliptical galaxy, leading to the 
additional requirement of a blue continuum color in 
e.g. $V-R$, consistent with the expected starburst nature 
of young galaxies.  This makes the LBG technique insensitive to 
heavily dust-reddened or evolved stellar populations.  

The selected population of galaxies is described in detail 
by \citet{giavalisco02} and \citet{steideletal03}.  
Star formation rates range from 10-1000 M$_\odot$ yr$^{-1}$ 
with a median value of $\sim 50$ M$_\odot$ yr$^{-1}$ 
after correction for reddening values ranging over 
$0\la$E($B-V$)$\la0.4$ \citep{pettini04}.    
Inferred stellar masses range over 
$6\times10^8$M$_\odot \la $ M$_* \la 10^{11}$M$_\odot$ with 
median value $2\times10^{10}$M$_\odot$.  
Implied stellar ages range over 
1 Myr$ \la t_* \la$ 2 Gyr with median age 500 Myr 
\citep{shapleyetal05}.  
Observed qualities of LBGs are summarized in Tables \ref{tab:z3} and 
\ref{tab:cq} below, giving values for the 
space density, clustering length and 
dark matter halo masses from \citet{adelbergeretal05a}, 
the SFR and stellar 
mass per object and stellar mass density from \citet{shapleyetal01}
and the cosmic SFRD from \citet{steideletal99}.

\subsection{Lyman Alpha Emitters (LAEs)}
\label{sec:lae}

Starbursting galaxies can emit 
most of their ultraviolet 
luminosity in the Lyman $\alpha$ line.  Because 
Lyman $\alpha$ photons are resonantly scattered in neutral hydrogen, 
even a small amount of dust will quench this emission.  Hence, selecting 
objects with strong Lyman $\alpha$ emission lines is expected to 
reveal a set of objects in the early phases of rapid star formation.  
These could either be young objects in their first burst 
of star formation or evolved galaxies undergoing a starburst due to 
a recent merger.  Selecting galaxies with strong emission lines also 
allows us to probe the high-redshift luminosity function dimmer than 
the  
``spectroscopic'' continuum limit of magnitude $R=25.5$ that is 
used to select the Steidel et al. LBG samples, since continuum 
detection is not necessary for spectroscopic confirmation 
using the emission line.  

Observed qualities of the Lyman Alpha Emitting galaxies (LAEs)
 are summarized in Tables \ref{tab:z3} and 
\ref{tab:cq} below, giving values for the 
SFR per object from \citet*{huetal98} and the 
space density, SFRD,  
clustering length and   
dark matter halo masses from MUSYC 
\citep{gawiseretal05b}.  

\subsection{Distant Red Galaxies (DRGs)}
\label{sec:drg}

The inability of the Lyman break selection technique to find 
intrinsically red objects can be overcome by using 
observed NIR imaging to select high-redshift galaxies 
via their rest-frame Balmer/4000\AA\ break.  Looking 
for a continuum break in $J-K$ selects objects 
at $2<z<4$, labelled Distant Red Galaxies (DRGs)
 \citep{franxetal03, vandokkumetal03}.   
\citet{reddyetal05} offer a comparison of the redshift 
distributions of objects selected by LBG/star-forming 
colors, DRGs selected in $J-K$, and the passive evolution 
and star-forming samples selected through 
their $BzK$ colors by \citet{daddietal04b}.  
Note that this comparison is somewhat biased as the spectroscopic 
follow-up was performed on a sample originally selected only by the 
LBG/star-forming criteria.  \citet{vandokkumetal05} report MUSYC 
results for an analogous comparison derived from a $K$-selected 
sample with inferred stellar masses $>10^{11}$M$_\odot$.  

Observed qualities of DRGs are summarized in Tables \ref{tab:z3} and 
\ref{tab:cq} below, giving values for the 
SFR and stellar mass per object from \citet{vandokkumetal04} and for the 
space density, SFRD, stellar mass density, clustering length and   
dark matter halo masses from MUSYC (Gawiser et al, in preparation).  

\subsection{Sub-Millimeter Galaxies (SMGs)}
\label{sec:smg}

The Sub-millimeter galaxies (SMGs) are selected using sub-millimeter 
bolometer arrays, e.g. SCUBA or MAMBO, which have poor spatial 
resolution, $\sim15''$.  Complementary high-resolution 
radio imaging is needed to obtain positions accurate enough to find 
optical counterparts or perform spectroscopy.  This means that the 
SMGs with redshifts are really jointly selected in both sub-mm 
and radio.  
Observed qualities of SMGs are summarized in Tables \ref{tab:z3} and 
\ref{tab:cq} below, giving values for the 
space density from \citet{chapmanetal03a}, the SFR per object and 
SFR density from \citet{chapmanetal05}, the 
clustering length from \citet{webbetal03} and the 
dark matter halo masses from MUSYC (Gawiser et al., in preparation). 

\subsection{Damped Lyman $\alpha$ Absorption Systems (DLAs)}
\label{sec:dla}

 The Damped Lyman $\alpha$ Absorption systems (DLAs) 
were introduced above in \S \ref{sec:HI}~  
Observed qualities of DLAs are summarized in Tables \ref{tab:z3} and 
\ref{tab:cq} below, giving the range 
of SFR per object for the two DLAs for 
which this quantity has been determined 
\citep[][see \citealp{wolfegp05} for a review]{molleretal02, bunker04}. 
Also shown are the 
SFR density from \citet{wolfegp03} and the  
clustering length and dark matter halo 
masses determined by \citet{cookeetal05b}.

\section{Clustering of protogalaxies}
\label{sec:clustering}

It seems appropriate to provide a brief summary of the method used 
to generate the clustering lengths and inferred dark matter halo 
masses given in the Tables.  
The spatial correlation function $\xi(r)=(r/r_0)^{-\gamma}$ 
is inferred 
by fitting a power-law to either the observed spatial or angular 
correlation function
of the sample.  If only angular positions are observed, the redshift 
distribution $N(z)$ must be measured spectroscopically and used to 
invert the Limber equation as described in \citet{giavaliscoetal98}.  
The Landy-Szalay estimator is typically used to estimate the angular 
or spatial correlation function of the datapoints and to correct for the 
so-called ``integral constraint'' caused by measuring the mean density 
of the population from the observed survey volume \citep{landys93}.  
The LBG, LAE, and DRG samples are large enough to use the 
correlation length $r_0$ measured from the 
auto-correlation function to determine the bias 
factor e.g. $b_{LBG}$, following  
\begin{equation}
\xi_{LBG-LBG}(r) = (r/r_0)^{-\gamma} =b^2_{LBG} \xi_{DM}(r), \; \; \;
\label{eq:auto}
\end{equation}
where $\xi_{DM}(r)$ is the dark matter autocorrelation function 
predicted by the $\Lambda$CDM cosmology.  
The SMG and DLA samples are small, so their 
cross-correlation with the more numerous LBGs is used 
to determine their bias
factor, e.g. 
\begin{equation}
\xi_{DLA-LBG}(r) =(r/r_0)^{-\gamma} = b_{DLA}b_{LBG} \xi_{DM}(r) . \; \; \;
\label{eq:cross}
\end{equation}

The bias factor
of each family of protogalaxies determines its 
typical dark matter halo mass
following
the method of \citet{mow96}, whose application to the 
cross-correlation function was first suggested 
by \citet{gawiseretal01}.  
This method also allows one to predict the number abundance of 
dark matter halos with mass above 
the given threshold mass and to compare this with the observed number 
density of the population to infer the average halo occupation number.

\begin{table}[!ht]
\caption{The $z=3$ universe.  References for entries are given in the 
text, with a few entries still to be determined from MUSYC, ALMA, and 
JWST.  Typical systematic uncertainties are a factor of two.}
\label{tab:z3}
\smallskip
\begin{center}
{\small 
\begin{tabular}{ccccc}
\tableline
\noalign{\smallskip}
TLA&Space density&SFR per object&Stellar mass per object&Clustering length ($r_0$)\\
   &[$h_{70}^3$ Mpc$^{-3}$]&[M$_\odot$yr$^{-1}$]&[M$_\odot$]&[h$_{70}^{-1}$ Mpc]\\
\noalign{\smallskip}
\tableline
LBG&$2\times10^{-3}$       &30               &$10^{10}$       &$6\pm1$\\  
LAE&$3\times10^{-4}$       &6               &MUSYC               &$4\pm1$\\
DRG&$3\times10^{-4}$       &200              &$2\times10^{11}$&$9\pm2$\\
SMG&$2\times10^{-6}$       &1000             &MUSYC               &$16\pm7$\\
DLA&ALMA                   &1--50            &JWST            &$4\pm2$\\
\noalign{\smallskip}
\tableline
\end{tabular}
}
\end{center}
\end{table}

\begin{table}[!ht]
\caption{Cosmological quantities.  References for entries are given in the 
text, with a few entries still to be determined from MUSYC and JWST.  Typical
systematic uncertainties are a factor of three.}
\label{tab:cq}
\smallskip
\begin{center}
{\small 
\begin{tabular}{cccc}
\tableline
\noalign{\smallskip}
TLA&SFR density&Stellar mass density&Dark matter halo mass\\
   &[M$_\odot$yr$^{-1}$$h_{70}^3$ Mpc$^{-3}$]&[M$_\odot$$h_{70}^3$ Mpc$^{-3}$]&[M$_\odot$]\\
\noalign{\smallskip}
\tableline
LBG&0.1               &$10^7$        &$3\times10^{11}$      \\  
LAE&0.002              &MUSYC         &$10^{11}$      \\ 
DRG&0.06             &$6\times10^7$  &$3\times10^{12}$      \\ 
SMG&0.02             &MUSYC          &$10^{13}$      \\ 
DLA&0.03            &JWST            &$10^{11}$      \\ 
\noalign{\smallskip}
\tableline
\end{tabular}
}
\end{center}
\end{table}

\section{Five Unsolved Problems in Galaxy Formation}
\label{sec:problems}

1. {\it What does a protogalaxy look like?}  
The term protogalaxy has 
been used loosely here and in the literature 
to describe young galaxies at high redshift.  Part of the difficulty 
is that once an object has sufficient stars to be observed in 
rest-frame UV or optical radiation, we consider it a galaxy.    
But before this time it is either unobservable or only observable 
in absorption (e.g. DLAs), X-ray emission from a supermassive 
black hole (quasars/AGN), 
or in far-infrared radiation from dust which could be 
enshrouding either a powerful AGN or rapid star formation.  If dark 
matter halo collapse, initial star 
formation and supermassive black hole formation all occur simultaneously, 
the formation epoch of the galaxy is well-defined, and the picture is 
simple.  But it is possible that many collapsed halos remain 
quiescent clouds of neutral gas until star formation is triggered by 
later mergers; these objects could comprise the half of DLAs that fail to show 
significant cooling in the [CII] 158 micron line 
\citep{wolfeetal04}.  The distribution of 
lag times between dark matter halo 
collapse, supermassive black hole formation, and rapid star formation 
remain uncertain.   

2.  {\it When/how did each component of the Galaxy form?}
Observations indicate that the thin disk formed at $z\simeq2$, but 
simulations have trouble creating disk galaxies.  One area now receiving 
attention is the manner in which 
      angular momentum coupling between dark matter and baryons 
affects bar/disk formation and the cuspiness of bulges.  
It is still not clear if the globular clusters should be considered Galactic 
components or were all formed earlier and captured, despite evidence that 
some globular clusters are captured dwarf galaxies.  Could globular 
clusters have formed in the same low-mass halos that met the Jeans 
threshold for collapse after recombination 
and hosted the Population III stars?  

3.  {\it When/how did galaxy sequences evolve?}
HST observations of morphologies of galaxies at $z>2$ imply that the 
Hubble sequence was not yet present.  This is somewhat subtle, as 
cosmological surface brightness dimming would make a face-on spiral 
appear very different at high redshift, but most 
objects display irregular morphology and even the most promising 
edge-on disk candidates show spectroscopic kinematics inconsistent 
with the presence of disks \citep{erbetal04}.  However, in the 
low-redshift ($z<1$) universe we see a clear bimodality in the 
distribution of galaxy properties, the so-called red and blue 
sequences \citep[e.g.][]{belletal04,kannappan04}.  Such bimodalities 
are unlikely to arise from cosmological 
structure formation but are presumably 
caused by baryonic physics and appear directly linked to the 
``downsizing'' behavior discussed in \S \ref{sec:initial}

4.  {\it What role did feedback play?}
The non-linear baryonic physics of star formation leads to highly 
energetic processes (ultraviolet radiation, stellar winds, 
supernova explosions) 
that can ionize or expel neutral gas that would 
otherwise participate in further star formation.  
 It is now clear that the processes of galaxy and 
supermassive black hole formation are 
intimately connected, as evidenced by striking correlations between 
the masses of black holes and the velocity dispersions (or masses) 
of bulges 
in which they 
are embedded 
\citep{gebhardtetal00, ferraresem00, kormendyr95, magorrianetal98}.  
Possible explanations include 
simultaneous hierarchical growth of galaxies 
and their central black holes through mergers 
\citep{haehneltk00,dimatteoetal05},   
a strong coupling between black hole accretion and star formation 
in proto-disks at high redshift \citep[e.g.][]{burkerts01}, 
and the effects of AGN feedback on the surrounding 
intergalactic medium  
\citep*{scannapiecoetal05}.  One way or another, it appears that 
feedback from AGN, supernovae, and galactic winds must regulate the 
joint formation of the bulge and central black hole.  Feedback may 
also play a role in determining the cuspiness of the dark matter 
halos, which does not appear consistent with profiles predicted from 
N-body simulations \citep{silk04}.  The galactic winds play a critical role in 
metal enrichment of the intergalactic medium and probably play 
a lesser role in ejecting neutral gas from the galaxies.  As mentioned 
above, supernova feedback may explain the apparent minimum 
galaxy mass. 

5.  {\it When/how was the universe reionized?}
A major area of ongoing investigation 
is the reionization epoch when the 
intergalactic medium was ionized.  Slightly inconsistent results have 
been reported for 
the reionization redshift from WMAP observations of the temperature-E-mode 
cross-power-spectrum ($z_r = 20\pm9$, \citealp{bennettetal03}) 
and the apparent end of 
reionization where the neutral hydrogen fraction dropped to 0.01 as 
seen in SDSS quasar spectra at $z\simeq 6.3$ 
\citep{fanetal02}.  It seems premature to  
hypothesize bimodal models of reionization 
where separate
 families of sources produce the ``early'' ionization seen by WMAP 
and the completion of reionization seen by SDSS.  Nonetheless, it is 
unclear at present which sources reionized the universe, and the 
leading candidates are the first generation of zero-metallicity 
stars (Population III) and starbursting galaxies including LBGs 
and LAEs.  The quasars have very hard, ionizing spectra but were not 
numerous enough to reionize the universe at $z>6$; they appear to 
dominate HeII reionization at $z\sim3$.  Significant uncertainties 
exist regarding the nature of the Population III stars:  did they 
form in $10^6$M$_\odot$ dark matter halos that collapsed after 
recombination, or in larger galaxies later on?  A top-heavy initial 
mass function (IMF) is presumed for Population III, but what was 
the exact 
mass range and nature of stellar death?  
Did multiple stars 
occur per halo, or did the death of the first very massive star prevent 
further star formation or cause sufficient metal enrichment to generate 
Population II stars?  


\section{Conclusions:  Coming Attractions}
\label{sec:attractions}

The speakers have been asked to discuss major advances expected in 
the coming decade.  For galaxy formation, I will go on record 
with three promising predictions and one slightly 
fascetious warning.  


{\it The coming years will see the 
unification of galaxy formation and evolution}.  
Until very recently, galaxy formation was studied at $z>2.5$ and 
galaxy evolution was studied at $z<1$ and the period $1<z<2.5$ was 
referred to as the ``redshift desert''.  But technological advances in 
NIR imaging and spectroscopy have made the rest-frame Balmer/4000\AA\ 
break and nearby emission lines available for study in distant galaxies.  
Development of these ``needle-in-a-haystack'' techniques now allows 
us to successfully find evolved galaxies at $z>2$ even though these 
objects may be rare at those epochs.  Hence we are beginning to study objects 
at $z\sim3$ that formed at $z>6$ which may turn out to be much easier than 
observing $z>6$ galaxies directly.  
Imaging with the Spitzer satellite is enabling the first studies of 
rest-frame near-infrared emission from $z>2$ galaxies, 
breaking degeneracies between age and dust.  
Deep imaging and slitless spectroscopy with the GALEX 
satellite are revealing the analogs of Lyman break galaxies 
at low redshift \citep{burgarellaetal05}.
These combined studies may make it possible 
to piece together a rough evolutionary sequence, 
e.g. DLA$\rightarrow$LAE$\rightarrow$LBG$\rightarrow$SMG$\rightarrow$DRG, 
that would form  
part of a {\it grand unified} model of high-redshift galaxies and AGN.  

{\it We will be able to study the interstellar medium
 in emission at high-redshift}.  
ALMA will enable studies of molecular gas in young galaxies through high-order 
CO lines. 
The [CII] 158 micron line, which dominates the cooling of the 
Cold Neutral Medium phase at both low and high redshift, should be detectable 
for galaxies with large gas mass or rapid cooling equilibrating the 
 heating due to starbursts.  
The current set of Early Universe Molecular Emission Line Galaxies 
consist mostly of quasars and 
are reviewed (and assigned the questionable TLA ``EMG'') 
by \citet{solomonv05}.  
Both CO and [CII] have now been detected in $z>6$ SDSS quasars, where they 
provide the best direct probes of the quasar host galaxies
\citep{bertoldietal03,walteretal04,maiolinoetal05}.  
Detecting these lines and the sub-millimeter dust continuum from 
protogalaxies with ALMA will allow us to probe 
a multivariate mass function of gas mass, molecular mass, dust mass, 
and stellar mass.  Even ALMA sensitivity may only allow detections 
of the tip of the gas-mass function, but this will provide a 
complementary 
set of objects to the tip of the rest-frame-UV and rest-frame-optical 
luminosity functions 
currently studied at high redshift, 
and much can be learned from the intersection 
and union of these samples.  

{\it High-redshift galaxies will be used to constrain dark energy properties}.   It has recently been shown 
\citep{seoe03,linder03,blakeg03} 
that the scale of baryon acoustic oscillations provides a ``standard 
rod'' that can be measured in the clustering of 
high-redshift galaxies.  
 The measurement will constrain the 
dark energy equation-of-state as a function of redshift, $w(z)$,  
via its influence on the expansion history 
of the universe.  The measurement can be performed at any redshift 
where the line-of-sight starting at $z=0$ is sufficiently influenced 
by the dark energy, making $z=1$ and $z=3$ equally acceptable.  
Of order a million redshifts are needed, and the most likely surveys 
to accomplish this ambitious goal appear to be HETDEX using the 
VIRUS instrument under construction for 
HET and the wide-field multi-fiber spectrograph 
KAOS proposed for Gemini.

{\it The rapidly increasing sophistication of studies of the 
high-redshift universe will generate even more jargon}.  We are 
already debating proper nomenclature for special categories of 
DLAs at lower column density (sub-DLAs) and those found in 
gamma-ray burst afterglows (burst-DLAs or bDLAs).  Four-letter 
object acronyms (FLOAs?) are going to be part of the future.    


\acknowledgements 
In terms of organization, comraderie, talks, and facilities, 
the 2005 Bash Symposium was a 5$\sigma$ event, which is inconsistent with 
gaussian random initial conditions, thereby proving ``intelligent 
design'' by the organizing committees.  
I thank the organizers for inviting me to
speak on my favorite topic and the editors for their hard work 
assembling this volume.  I acknowledge valuable conversations with 
Pieter van Dokkum, Priya Natarajan, Jason Tumlinson and Meg Urry
 while outlining this talk.  I thank the MUSYC Collaboration for allowing 
me to show results in preparation.    
This material is based upon work supported by the National Science 
Foundation under Grant. No. AST-0201667, 
an NSF Astronomy and Astrophysics Postdoctoral 
Fellowship (AAPF) awarded to E.G.



\end{document}